\documentclass{aa}
\usepackage{graphicx}
\usepackage{natbib}
\bibpunct{(}{)}{;}{a}{}{,}

\newcommand{\SiIII}{\ion{Si}{iii}}
\newcommand{\CI}{\ion{C}{i}}

\newcommand{\CaX}{Ca\ts$\scriptstyle{\rm X}$}

\newcommand{\FeXV}{Fe\ts$\scriptstyle{\rm XV}$}
\newcommand{\OVI}{O\ts$\scriptstyle{\rm VI}$}
\newcommand{\CaII}{Ca\ts$\scriptstyle{\rm II}$}
\newcommand{\phsm}{photons~s$^{-1}$~m$^{-2}$~sr$^{-1}$}
\newcommand{\HH}{H$_2$}

\usepackage{txfonts}
\begin{document}

   \title{SUMER-Hinode observations of microflares: excitation of
   molecular hydrogen}

   \author{D. E. Innes}

   \offprints{D. E. Innes}

   \institute{Max-Planck Institut f\"{u}r Sonnensystemforschung,
  37191 Katlenburg-Lindau, Germany\\
              \email{innes@mps.mpg.de}
             }

  \date{Received ----; accepted ----}

\abstract
{Concentrations of \HH\ have been detected by SUMER in
active region plage. The \HH\ is excited by \OVI\ line emission at 1031.94~\AA\ which, although not
observed, must be brightening along with the observed  transition
region line, \SiIII\ 1113.24~\AA.}
{We investigate the excitation of \HH\ and demonstrate the association
between the observed \HH\ emission and footpoints of X-ray microflares.}
{We have made co-ordinated observations of active region plage with the spectrometer
SUMER/SoHO in lines of \HH\ 1119.10~\AA\ and \SiIII\ 1113.24~\AA\ and with
XRT/Hinode
X-ray and SOT/Hinode \CaII\ filters.}
{In six hours of observation, six of the
seven
\HH\ events seen occurred near a  footpoint of a brightening
X-ray loop. The seventh is associated with an unusually strong \SiIII\ plasma
outflow.}
{Microflare energy dissipation heats the chromosphere, reducing its
 opacity, so that \OVI\ microflare
emission is able to reach the lower layers of the chromosphere and
excite the \HH.}


\titlerunning{Excitation of \HH}
 \maketitle


\section{Introduction}

Solar \HH\ emission is strong in ultraviolet spectra of
sunspots and has also been seen in flares \citep{Bartoe79}. In
 the quiet Sun it is present but extremely weak \citep{Sandlin86}.
Here we report the first observations of \HH\ concentrations in bright
active region plage.
The observed \HH\ line at 1119.10 \AA\ is the $1-3$ transition in the Werner
series, excited by \OVI~1032~\AA\
\citep{Bartoe79, Schuehle99}:

\begin{eqnarray}
\mathrm{H}_2 (v\arcsec=1, \mathrm{X}^1\Sigma^+_\mathrm{g}) + h\nu
(\mathrm{O\ VI})
& \longrightarrow & \mathrm{H}_2 (v\arcmin=1, \mathrm{C}^1\Pi_\mathrm{u})
\longrightarrow {}
\nonumber\\
& & {} \longrightarrow \mathrm{H}_2 (v\arcsec=3, \mathrm{X}^1\Sigma^+_\mathrm{g}
). \nonumber
\end{eqnarray}
The 1119.10~\AA\ line is about $60~\%$ as bright as the strongest $1-4$ Werner
line at 1164~\AA. The \HH\ is believed to be formed just above the temperature minimum at
around 4200~K.
Its strength is expected to correlate with the
\OVI\ intensity, as well as with the chromosphere structure in and above
the \HH\ region \citep{Jordan78}.

In this letter, three types of \HH\ plage events are discussed.
The strongest coincided with ribbon-like \CaII\ chromospheric brightening at
a footpoint of an X-ray microflare.
The second occurred near the footpoint of a brightening X-ray loop
with no signature in \CaII, and the third had
neither X-ray emission nor a \CaII\ signature but very strong
transition region outflow.
All three events highlighted here occurred in three hours
on one of the observing days.
During the second day of observation,
three \HH\ events were detected in three hours,
and all were associated with X-ray loop brightening
with no \CaII\ signature.

\section{Observations}   
\begin{figure}
\centering
\includegraphics[width=8 cm]{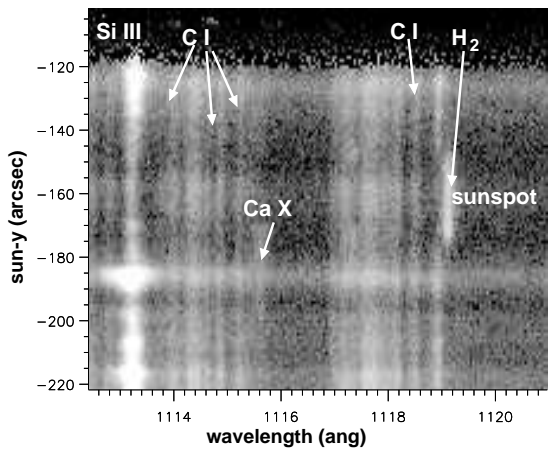}
\caption{SUMER spectrum of the region across and around the sunspot,
taken on 29 Apr 2007 at 02:25 UT.}
\label{spectrum}
\end{figure}

Hinode \citep{Kosugi07} and SUMER \citep{Wetal95} observed  a small active region (AR
10953) on 29 and 30 Apr 2007. The region had produced several B and one
C class flare four days earlier. All
events during the observing period discussed here were below B class.
On each day, SUMER made six rasters across the plage and sunspot. Each raster
took 30 min. The lines observed were \HH\ 1119.10~\AA\ ($4.2\times10^3$~K),
\CI\ multiplet at 1114.39~\AA\ and 1118.41~\AA\ ($10^4$~K), \SiIII\ 1113.24~\AA\
($6\times10^4$~K), and \CaX\ 2x557~\AA\ ($7\times10^5$~K), where the approximate
formation temperatures of the lines are given in brackets.
 The spectrum across the sunspot is shown in Fig.~\ref{spectrum}.
The sunspot is seen predominantly in the  \HH\ line.

Hinode made simultaneous observations with the X-Ray Telescope
\citep[XRT;][]{Golub07}
through both the Ti-poly and
Al-thick filters with 1.5~min cadence.
Observations through  both the \CaII\ and G-band filters
were made with the Broadband Filter Imager
of the Solar Optical Telescope
\citep[SOT;][]{Tsuneta07}
with a 1~min cadence. The Extreme ultraviolet Imaging Spectrometer
\citep[EIS;][]{Culhane07} observed \FeXV\ 284.25~\AA\ ($2\times10^6$~K),
using the wide 266\arcsec\ slit and
15~s cadence. To obtain the coalignment between the X-ray images and SUMER,
the EIS images were very useful because they bridged the gap between
SOT \CaII\ and XRT. In particular, during the strongest microflare
event, EIS captured both the X-ray loop seen with XRT and the ribbon-like footpoint
structure seen in the
SOT \CaII\ images. This provided the SOT, XRT, and EIS coalignment.
The SUMER-SOT coalignment was made by comparing SOT
G-band and SUMER \HH\ images. Additional checks were done by comparing SUMER
\SiIII\ and EIS \FeXV. The coalignment between SUMER and XRT
is believed to be better than 5\arcsec, with greater accuracy in the
north-south (sun-y) direction.

\begin{figure*}
\centering
\includegraphics[width=14 cm]{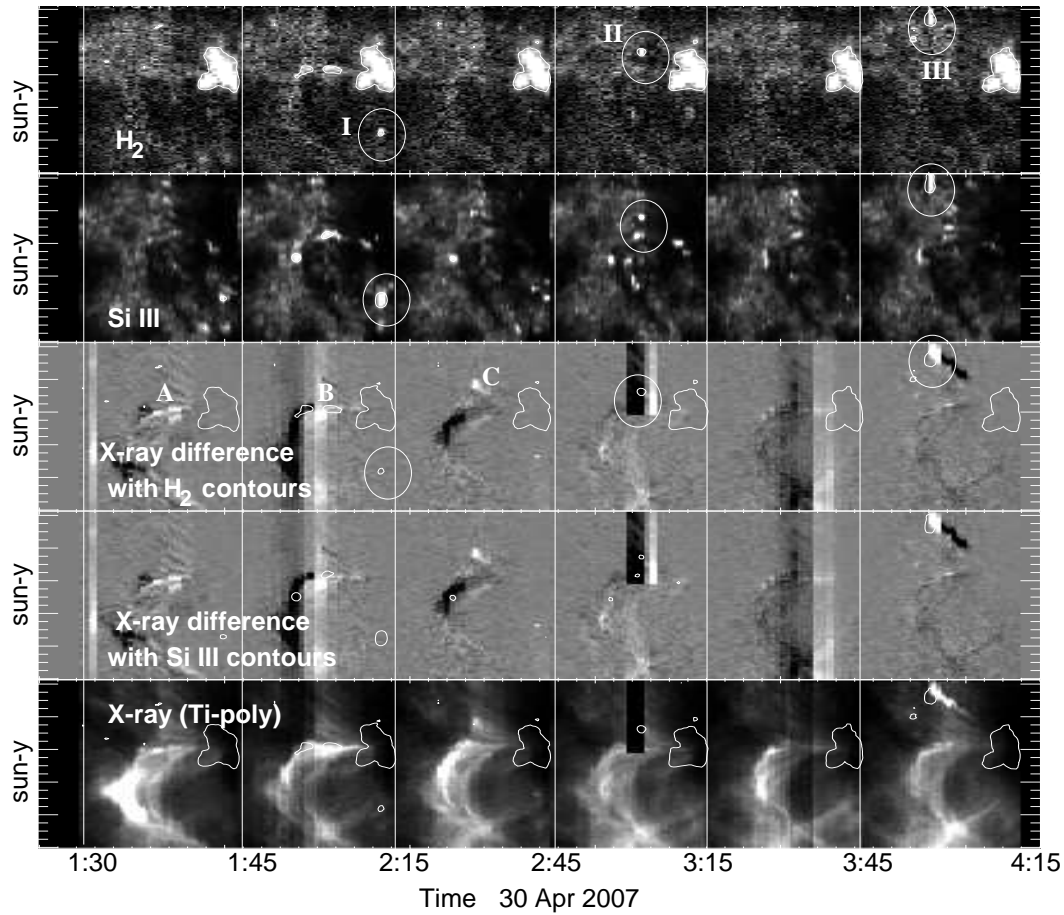}
\caption{The six raster scans taken in \HH\ and  \SiIII, and the equivalent
X-ray time-slice rasters. The X-ray rasters are constructed by stacking, for each
SUMER observation,
the cospatial XRT slices
from the XRT Ti-poly images closest in time. The difference images are
computed by subtracting the
preceding XRT Ti-poly image. The plage \HH\ brightenings
(labelled I, II, III)
are circled in the \HH, \SiIII, and top X-ray difference raster. The \HH\ contours
are at $8\times10^{-3}$~\phsm, and \SiIII\
contours are at 1.2~\phsm. All images have a linear intensity scale.
The sun-x and sun-y co-ordinates are as in Fig.~\ref{xrays}}
\label{rasters}
\end{figure*}

\begin{figure}
\centering
\includegraphics[width=8 cm]{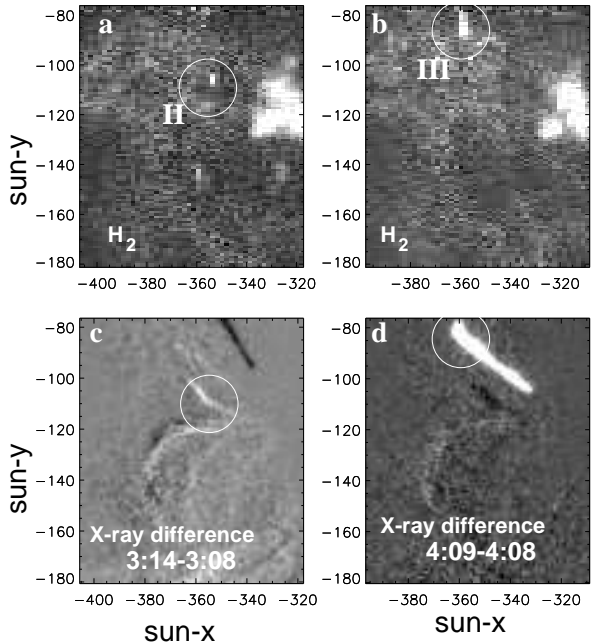}
\caption{SUMER raster images of \HH\ (top row) and, below, the XRT Ti-poly
difference images at
the time of the \HH\ brightenings.}
\label{xrays}
\end{figure}

The six rasters taken on 30 Apr 2007 are shown as a time series in
Fig.~\ref{rasters}. Here, the sunspot is the bright \HH\ region on the right of
each raster. The linear intensity scale used to display the images
accentuates the few plage brightenings, both in the \HH\ and the \SiIII\
images.
The three brightest \HH\ concentrations are
circled and labelled I, II, III.
Each \HH\ concentration coincides with intense \SiIII, but the inverse is
not true. Not all \SiIII\ brightenings are associated with \HH.
The relationship between
the \HH\ and X-ray brightenings is shown in the third row of Fig.~\ref{rasters}.
 The two strongest \HH\ events,
II and III, are associated with X-ray brightenings. Unfortunately,
the X-ray brightening at II is exaggerated by a data gap.
Its brightening is seen better in
the XRT difference images displayed in Fig.~\ref{xrays}.
Event III was much the brightest X-ray event during the observing period.
This is the event, mentioned previously, that coincided with a small, bright
ribbon-like structure seen in \CaII\  images.
Close inspection of the X-ray difference rasters reveals additional X-ray
brightenings, labelled A, B, and C in the third row of Fig.~\ref{rasters}.
Event B is associated with weak \HH\ and \SiIII, and A and C only with weak
\SiIII.

\begin{figure*}
\centering
\includegraphics[width=12 cm]{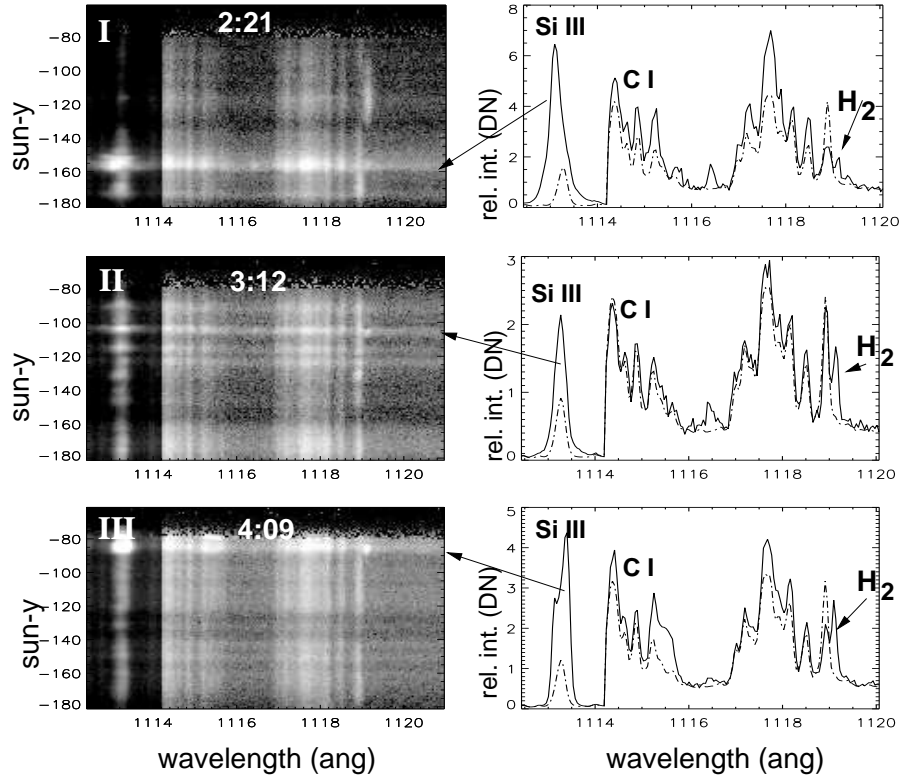}
\caption{SUMER spectral images and profiles of the \HH\ brightenings. The \SiIII\
intensity is reduced by a factor 20 compared to the rest of the spectrum, so that
they can be displayed with the same scale.
Each \HH\ spectrum is compared to the average plage spectrum, which has
been scaled so that their continua at 1116~\AA\ match.}
\label{profiles}
\end{figure*}

The SUMER spectra of the circled events (I, II, and III) are shown in
 Fig.~\ref{profiles}. There is no simple relationship
between the \SiIII\ strength and the \HH. In each case the \HH\ strength is
$1-1.5$ DN, whereas the \SiIII\ varies by a factor of three. It is
noticeable that the \SiIII\ emission is generally broader. In event~II, it is
offset by $2\arcsec$ from the \HH.

\section{Discussion}
The \HH\ intensity depends on both the strength of the \OVI\ and the
atmospheric opacity between the \OVI\ and \HH\ layers. If the \OVI\ follows
the \SiIII\ brightness, then an increase in transition region emission cannot
be the only reason for enhanced \HH\ excitation, because there are several examples
of bright \SiIII\ and no observed \HH. The analyses by \citet{Jordan78} and \citet{Bartoe79}
of \HH\ in a sunspot and quiet Sun led them to conclude that the opacity over
the sunspot is about an order of magnitude less than in the quiet Sun. This
suggests a similar explanation for the \HH\ concentrations in bright plage
because energy dissipation at X-ray loop footpoints is commonly associated with
chromospheric evaporation, and the two strongest events discussed here
and all three
events seen on  29 April 2007 were associated with
X-ray loop brightening. The one event not associated with X-ray
brightening showed significant plasma outflow from the transition region.

If chromospheric evaporation due to high-energy particles
accelerated in the microflare reduces the opacity of the chromosphere,
one might expect to find a relationship between the X-ray emission and the
\HH. This is not confirmed by the observations presented here, but this could
be because the observed \HH\ strength depends critically
on the position of the loop footpoint in relation to the SUMER field-of-view
at the time of the brightening. The reason more \SiIII\ and X-ray
coincidences occur is probably because \SiIII\ brightenings come from higher
up in the atmosphere and cover a larger area, and not just from the loop footpoint.

Future observations with SUMER will measure the \OVI\ 1032~\AA\ and \HH\
intensities almost simultaneously. This, together with extended SOT observations
 and analysis of the chromospheric dynamics,
  will help determine the extent and influence of opacity changes on the \HH\
  intensity.

\begin{acknowledgements}
I would like to thank Maria Madjarska for her constructive comments and
careful reading of this letter.
Hinode is a Japanese mission developed and launched by ISAS/JAXA,
collaborating with NAOJ, as domestic partner, and NASA (USA) and STFC (UK)
as international partners. Scientific operation of the Hinode mission is
conducted by the Hinode science team organized at ISAS/JAXA.
Support for the postlaunch operation is provided by JAXA and NAOJ, STFC,
NASA, ESA (European Space Agency), and NSC (Norway).
SUMER is financially supported by DLR, CNES, NASA, and the ESA PRODEX program
(Swiss contribution). SoHO is a project of international co-operation between
ESA and NASA.
We are grateful to all teams for their efforts in the design, building, and
operation of the Hinode and SoHO missions.
\end{acknowledgements}

\bibliographystyle{aa}

\end{document}